\documentclass[prd,aps,showpacs,showkeys,twocolumn,10pt,nofootinbib,altaffilletter]{revtex4-1}
\usepackage{amssymb,amsmath,amsthm}
\usepackage{mathrsfs}
\usepackage[applemac]{inputenc}

\usepackage{xcolor}
\usepackage{graphicx}
\usepackage{dcolumn}
\usepackage{bm}
\usepackage{hyperref}

\hypersetup{
    colorlinks=true,
    linkcolor=black,
    citecolor=black,
    filecolor=black,
    urlcolor=black,
}

\begin{document}


\title{The classical and quantum fate of the Little Sibling of the Big Rip in $f(R)$ cosmology}

\author{Teodor Borislavov Vasilev}
\email{teodorbo@ucm.es}
 \affiliation{Departamento de F\'isica Te\'orica, Universidad Complutense de Madrid, \\E-28040 Madrid, Spain.}

\author{Mariam Bouhmadi-L\'opez}%
 \email{mariam.bouhmadi@ehu.eus}
\affiliation{%
 Department of Theoretical Physics, University of the Basque Country, UPV/EHU, P.O. Box 644, 48080 Bilbao, Spain
}
\affiliation{
IKERBASQUE, Basque Foundation for Science, 48011, Bilbao, Spain.
}%

\author{Prado Mart\'in-Moruno}
\email{pradomm@ucm.es}
\affiliation{
 Departamento de F\'isica Te\'orica and  IPARCOS, Universidad Complutense de Madrid, \\E-28040 Madrid, Spain.
}

\date{\today}

\begin{abstract}
The Little Sibling of the Big Rip is a cosmological abrupt event predicted by some phantom cosmological models that could describe our Universe.
When this event is approached the observable Universe and its expansion rate grow infinitely, but its cosmic derivative remains finite. 
In this work we have obtained the group of metric $f(R)$ theories of gravity that reproduce this classical cosmological background evolution.
Furthermore, we have considered the quantization of some of the resulting models in the framework of quantum geometrodynamics, showing that the DeWitt criterion can be satisfied.
Therefore, as it also happens in General Relativity, this event may be avoided in $f(R)$ quantum cosmology.
\end{abstract}


\maketitle


\section{Introduction}

What is the final fate of the Universe? This question can be addressed in a scientific context since the formulation of General Relativity (GR). Einstein's theory allows us to describe the gravitational physics of small systems, contained in our laboratory, and of the largest gravitational system, that is the Universe, getting through all the astrophysical scales. From another point of view, until the date GR has passed all the observational tests, from those in the weak field regime to those of the strong gravitational events that generated the gravitational waves recently measured by the LIGO-Virgo collaboration (first detection in \cite{GW}).

Nevertheless, we have also had an amazing surprise in the field of gravitation 20 years ago. That is the discovery that the expansion of the Universe is currently accelerating \cite{Riess,Perlmutter}. This discovery changed our understanding about how could be the Universe's future. We now know that it is not probable that the Universe will reach a big crunch singularity. This is because the description of the accelerated expansion of the Universe in the framework of GR requires the introduction of dark energy and, at least for the most common models that we have, this fluid will dilute slower than matter (if it does it). The standard model of cosmology assumes that dark energy (DE) is a cosmological constant. In this case the Universe will tend to be described by a de Sitter space and approach a thermal death, slightly different in nature than that predicted by decelerated models. However, if the expansion is faster than that predicted by a cosmological constant, that is known as super-accelerated expansion, the Universe could have a different fate. All the structures of the Universe and the Universe itself might be ripped apart at a big rip singularity \cite{Starobinsky:1999yw,Caldwell:2003vq}. The Universe could also reach a cosmic singularity characterized by a divergent rate of expansion but a finite size of the observable Universe, freezing its evolution at a big freeze \cite{BouhmadiLopez:2006fu,BouhmadiLopez:2007qb}. (See also \cite{Barrow1,Barrow2,libroT0ab,2019zvz}, and references therein, for other examples of cosmic singularities.) Whereas the big rip and big freeze would take place at a finite cosmic time, the cosmic catastrophe may also be delayed an infinite cosmic time, in which case the singularity is called an abrupt cosmic event. Indeed, the little rip is just a big rip that would take place at an infinite cosmic time, although the cosmic structures will be ripped apart at a finite time \cite{Frampton:2011sp} (see also, \cite{Ruzmaikina1970}). The Little Sibling of the big rip (LSBR) is another abrupt cosmic event. It is characterized by the divergence of the observable Universe and the expansion rate, keeping the derivative of this rate a finite value \cite{LSBR}. For observational constraints on this type of models see references \cite{Albarran:2016mdu, Albarran:2017kzf,Bouali}. Nonetheless, the common belief is that these singularities will be cured or avoided in the quantum realm, as it is assumed to happen with the big bang (see \cite{2019zvz, Nojiri:2005sx,Nojiri, Nojiri3} for reviews on the topic).
 In Table \ref{tab:singularities} we include a summary of these rip-like curvature singularities. 
Further information can be found in references \cite{Caldwell:1999ew,Frampton:2011sp,LSBR,Frampton:2011aa}.

\begin{table}\label{tab:singularities}
\begin{tabular}{|c|c|c|c|c|}
\noalign{\smallskip}\noalign{\smallskip}
\hline \\[-1em]
& \parbox[c]{1.8cm}{$t_{\rm rip}$}& $a$ &  $H$ & $\dot{H}$  \\
\hline
\hline
Big Rip & finite & $\infty$ & $\infty$ & $\infty$ \\ 
\hline
Little Rip & $\infty$ & $\infty$ & $\infty$ & $\infty$ \\
\hline
LSBR & $\infty$ & $\infty$ & $\infty$ & finite  \\
\hline
Pseudo-Rip & $\infty$ & $\infty$ & finite & finite  \\
\hline
\end{tabular}
\caption{Characterization of the rip-like events discussed in the introduction by means of the time of occurrence, the scale factor $a$, the Hubble parameter $H$, and its cosmic derivative $\dot{H}$.
 The Pseudo-Rip corresponds to a mild event (before which the structures are ripped apart) rather than to a curvature singularity.
 Note that in GR the divergence in $H$ implies that the energy density blows up.}
\end{table}

In GR that kind of super-accelerated expansion is modelled by a dark energy fluid of phantom nature. This phantom energy is characterized by an energy density that increases with time and may have associated some potential pathologies \cite{Caldwell:1999ew,Carroll:2003st} (for an effective phantom behaviour in $f(R)$ gravity, see for example  \cite{Motohashi:2010tb,Motohashi:2011wy, introfR, introfR 2, 2017ncd, fRQC}). 
On the other hand, alternative theories of gravity have attracted a huge interest over the past decades as a possible alternative framework  to describe the cosmic phases of accelerated expansion of the Universe.
 In particular, $f(R)$ metric theories of gravity are one of the simplest ways to build such an alternative framework, which can give an explanation for the observed cosmic acceleration without the need of dark energy. (For an introduction to $f(R)$ metric theories see, for example, references~\cite{introfR,introfR 2, introfR 3,2009hc,2011et,2017ncd, Rocco}.)
The observational data currently available can be used to constrain and set a selection rule among the existing theoretical models (some examples can be found in references~\cite{Okada,Dossett,delaCruz,Quasar}). Several reconstruction methods have been developed within $f(R)$ theories to select a particular theory that can describe an identical background cosmic evolution that a given general relativistic model but without introducing dark energy \cite{Recon,fRgravity,reconMGravity,NewreconfR,LCDMreconfR,ref1}. In this work we will investigate which $f(R)$ theories of gravity predict an accelerated expansion leading to a LSBR event, which is compatible with current observations. Up to our knowledge, this is the first study of $f(R)$ theories with a LSBR in the literature.

As the same background cosmic evolution can be described by GR or $f(R)$ gravity, one could wonder whether cosmic singularities will be avoided in the quantum realm for different underlying fundamental theories of gravity \cite{2019zvz}.  In the framework of quantum geometrodynamics several works have evaluated different kinds of cosmological singularities \cite{Dabrowski:2006dd,fateQsingularities,Albarran:2016ewi}. Furthermore, some works have also  investigated this issue for alternative theories of gravity by formulating a modified quantum geometrodynamical framework \cite{2016dcf,fRQC,2018tel,2018mpg}. (See also reference \cite{Ruf:2017bqx} for a different approach to quantum $f(R)$ gravity.) The avoidance of the LSBR in quantum geometrodynamics have been considered in references~\cite{Albarran,EoS DE 2}. In the present work we will consider the possibility of avoiding the LSBR in $f(R)$ quantum cosmology.

This paper is organized as follows: In Section \ref{sec:Cpart} we consider that the LSBR event could take place if gravity is described by a $f(R)$  theory. Thus, in the first place, we briefly review the characteristics of the LSBR in GR and the basics of the reconstruction method for metric $f(R)$  theories of gravity, in sections \ref{sec:LSBR} and \ref{sec:recmeth}, respectively. Then, in section \ref{sec:fRLSBR}, we apply the reconstruction method to describe the same GR dynamics that we have reviewed in \ref{sec:LSBR} with $f(R)$ theories. Thus, we obtain the group of metric $f(R)$ theories of gravity that predict a LSBR.
In Section \ref{sec:Qpart}, we study the LSBR in the framework of $f(R)$ quantum geometrodynamics. For that aim, we perform a brief summary of quantum geometrodynamics for an arbitrary $f(R)$ theory in section \ref{sec:MWD}. Then, in section \ref{sec:WDWfR}, we analyse the behaviour of the wave function of the Universe nearby the LSBR event. The analysis is made through the modified Wheeler-DeWitt (WDW) equation for the reconstructed $f(R)$ setup considering the DeWitt (DW) criterion. We summarise and present our conclusion in Section \ref{sec:Summary}. Finally, in Appendix \ref{WKBapp} we discuss several details about the WKB approximation carried to solve the modified WDW equation.


\section{\label{sec:Cpart} The LSBR in $f(R)$ classical cosmology}

The so-called ``reconstruction method'' is a technique used to recover a given background cosmological evolution in the framework of a family of alternative theories of gravity by restricting attention to a particular theory.
For example, in the framework of $f(R)$ theories of gravity, one can select the function $f(R)$ that allows us to reconstruct a given background cosmological evolution  \cite{Recon,fRgravity,reconMGravity,NewreconfR,LCDMreconfR,ref1}. In this section, we will apply this method to obtain the group of metric $f(R)$ theories of gravity leading to a LSBR abrupt cosmic event.

\subsection{The LSBR}\label{sec:LSBR}
Let us briefly summarize the phenomenology of the LSBR in GR. 
Homogeneous and isotropic cosmological solutions are described by a Friedmann-Lema\^itre-Robertson-Walker (FLRW) metric given by 
\begin{equation}\label{metric}
ds^2=-dt^2+a(t)^2ds^2_3 ,
\end{equation}
where we have set $8\pi G$ and $c$ equals to unity. The function $a(t)$ is the scale factor and $ds_3^2$ represents the $3$-dimensional metric, whose spatial curvature is not fixed at this point. 
Assuming that the Universe is filled with a perfect fluid, the Einstein equations reduce to the well-known Friedmann equations, cf. \cite{peacock},
\begin{align}\label{F.E.}
\frac{\dot{a}^2}{a^2}&=H^2=\frac{1}{3}\rho-\frac{k}{a^2}   ,\\
\frac{\ddot{a}}{a} &=\dot{H}+H^2=-\frac{1}{2}\left(p+\frac{\rho}{3}\right) \label{FE2}  ,
\end{align}
where the dot represents the derivative with respect to the cosmic time, $H$ stands for the Hubble rate, $k$ is the spatial curvature of the Universe, 
and $p$ and $\rho$ are the pressure and energy density of the fluid, respectively. 
According to the interpretation of the observational data in the framework of GR, that cosmic fluid is formed by dark energy, matter, and radiation.
We know that nowadays the fractional energy densities of matter (M) and dark energy (DE) are $\Omega_{M,0}\sim 0.306$ and $\Omega_{DE,0}\sim0.694$ \cite{Planck 1, Planck 2}, respectively. While the radiation contribution can be ignored at present.
So, the dominant cosmic ingredient today is dark energy, and it will be even more dominant in the future since matter tends to dilute (faster).
Therefore, from a practical point of view, we can neglect the contribution of the matter and radiation components to study the asymptotic evolution of these models.
So, we consider that $p$ and $\rho$ in equations (\ref{F.E.}) and (\ref{FE2}) are those corresponding to dark energy.

The LSBR is a cosmological event that takes place at an infinite cosmic time at which the Hubble rate and the scale factor blow up but the cosmic derivative of the Hubble rate does not.
It is obtained by assuming a dark energy equation of state that deviates slightly from that of a cosmological constant by a constant factor. This is
\begin{equation}\label{EoS}
p+\rho=-A/3  ,
\end{equation}
being $A$ a small positive parameter, see references~\cite{LSBR, EoS DE,EoS DE 2}. The conservation of the energy momentum tensor implies that $\rho$ evolves with the scale factor as 
\begin{equation}\label{rho(a)}
\rho= \Lambda+A\ln\frac{a}{a_0}   ,
\end{equation}
with $\Lambda$ an integration constant playing the role of an effective cosmological constant at present and $a_0$ representing the present scale factor of the Universe.
The equation of state parameter $w$ reads
\begin{equation}
w=\frac{p}{\rho}=-1-\frac{A}{3\left(\Lambda+A\ln \frac{a}{a_0}\right)}  .
\end{equation}
It should be noted that $w$ approaches the value $-1$  asymptotically as the scale factor evolves towards the future. However, the behaviour is not that of a de Sitter model since the energy density is not constant and it even tends to blow up at the LSBR.
As it was shown in reference~\cite{LSBR}, 
although the event takes place at infinite cosmic time in the future, the cosmological bounded structures are destroyed at a finite time scale. 
Furthermore, the evolution described by this model was shown to be compatible with that modelled by the $\Lambda$CDM scheme and constrained observationally in~\cite{Bouali}.


\subsection{The reconstruction method}\label{sec:recmeth}

We want to find a $f(R)$ theory of gravity that describes the same cosmic evolution as the model we have summarized and, therefore, predicts the occurrence of a LSBR event. With this aim, we follow a line of reasoning similar to that presented in reference~\cite{ref1} and note that the scalar curvature of the considered GR model satisfies the following relation
\begin{equation}\label{R(H)}
R=6\left(\dot{H}+2H^2+\frac{k}{a^2}\right)=\rho-3p  ,
\end{equation}
where in the last step we have used the Friedmann equations (\ref{F.E.}) and (\ref{FE2}).
Moreover, taking into account these Friedmann equations and the equation for the conservation of the stress energy tensor of the perfect fluid, that is
\begin{equation}
\dot{\rho}+3H(p+\rho)=0   ,
\end{equation}
one can obtain
\begin{align}\label{dot rho and p}
\dot{\rho}&=-3(p+\rho)\left(\frac{1}{3}\rho-\frac{k}{a^2}\right)^{\frac{1}{2}}   , \\
\dot{p}&=-3(p+\rho)\left(\frac{1}{3}\rho-\frac{k}{a^2}\right)^{\frac{1}{2}}\frac{dp}{d\rho}   ,
\end{align}
where, for the time being, we have just assumed $p=p(\rho)$. 
These two equations can be used to get
\begin{equation}\label{dot R}
\dot{R}=-3\,(p+\rho)\left(\frac{1}{3}\rho-\frac{k}{a^2}\right)^{\frac{1}{2}}\left(1-3\frac{dp}{d\rho}\right)  .
\end{equation}

On the other hand, the cosmic field equations will not be exactly the Friedmann equations for an alternative theory of gravity. Indeed, for a $f(R)$-theory of gravity, which is described by the action 
\begin{equation}
 S=\frac{1}{2}\int d^4x\sqrt{-g}f(R),
\end{equation}
the first modified Friedmann equation reads (see, for example, references~\cite{ref1,M.F.E. 1970, M.F.E. libro})
\begin{equation}\label{M.F.E.}
3H^2 \frac{df}{dR}=\frac{1}{2}\left(R\frac{df}{dR}-f\right)-3H\dot{R}\frac{d^2f}{dR^2	}-3\frac{k}{a^2}+\rho_m ,
\end{equation}
where $\rho_m $ is the energy density of the minimally coupled material content.
We are interested in a $f(R)$ theory with a background cosmological expansion equivalent to that provided by GR for a particular kind of fluid $p=p(\rho)$. 
So, in the next section, we will assume that $\rho_m=0$ and that the dark energy evolution is mimicked by the modifications appearing in the Friedmann equation due to $f(R)\neq R$.
Therefore, such a theory must be a solution to equation (\ref{M.F.E.}) that satisfies equations (\ref{F.E.}) to (\ref{dot R}), 
where $\rho$ and $p$ are now understood as the effective energy density and pressure that encapsulate the modifications with respect to GR  (see, for example, reference~\cite{ref1}).

\subsection{$f(R)$ theories predicting the LSBR}\label{sec:fRLSBR}
Let us now restrict our attention to a flat FLRW and an effective equation of state given by equation (\ref{EoS}). Therefore, the expressions for the Hubble rate,  the scalar curvature and its cosmic time derivative are given by
\begin{align}
H^2&=\frac{1}{3}\rho ,\\
R&=4\rho +A   ,\\
\dot{R}&=\frac{4\sqrt{3}}{3}A\sqrt{\rho}   ,
\end{align}
respectively. We emphasize that now $\rho$ and $p$ are effective quantities encapsulating the modifications of the predictions with respect to GR. Taking these expressions into account in equation (\ref{M.F.E.}), one obtains $f(R)$ as a function of $\rho$. Substituting then $\rho=(R-A)/4$, one gets
\begin{equation}
A(A-R)\frac{d^2}{dR^2}f(R)+\frac{1}{4}(R+A)\frac{d}{dR}f(R)-\frac{1}{2}f(R)=0  .
\end{equation}
Considering $y=\frac{R-A}{4A}$, this equation can be expressed as
\begin{equation}
y\frac{d^2}{dy^2}f-\left(\frac{1}{2}+y\right)\frac{d}{dy}f+2f=0   .
\end{equation}
The above expression is known as the Kummer's confluent hypergeometric equation, cf.~13.1.1 of reference~\cite{libroFunciones}. The general solution is
\begin{equation}
\begin{split}
f(R)&=\tilde{c}_1 \ {}_1F_1\left(-2;-\frac{1}{2};\frac{R-A}{4A}\right) \\
&\quad + \tilde{c}_2\left(\frac{R-A}{4A}\right)^{\frac{3}{2}} {}_1F_1\left(-\frac{1}{2};\frac{5}{2};\frac{R-A}{4A}\right)   ,
\end{split}
\end{equation}
where ${}_1F_1$ is the confluent hypergeometric function or Kummer's function, see references~\cite{Libro ODE, Riley,libroFunciones},
and $\tilde{c}_1$ and $\tilde{c}_2$ are arbitrary constants.
 An important feature of ${}_1F_1(a; b; y)$ is that it can be related with the generalised Laguerre polynomials  when $a$ is a negative integer but $b$ is not, cf.~table~13.6 of reference~\cite{libroFunciones}. 
 Hence,
\begin{equation}
\begin{split}
{}_1F_1&\left(-2;-\frac{1}{2};\frac{R-A}{4A}\right)\propto L_{2}^{-\frac{3}{2}}\left(\frac{R-A}{4A}\right) \\
&\quad \propto \left( A^2-6AR+R^2\right)  ,
\end{split}
\end{equation}
where in the last step we have made use of the Rodrigues' formula (see, for example, equation 22.1.6 of reference~\cite{libroFunciones}). Therefore, the general expression for $f(R)$ takes the form
\begin{equation}\label{f(R)}
\begin{split}
f(R)& = c_1\left(A^2-6AR+R^2\right) \\
&\quad+c_2 \left(\frac{R-A}{4A}\right)^{\frac{3}{2}} 
{}_1F_1\left(-\frac{1}{2};\frac{5}{2};\frac{R-A}{4A}\right)  ,
\end{split}
\end{equation}
with $c_1$ and $c_2$ being arbitrary constants.
We emphasize that the group of $f(R)$ metric theories given in equation (\ref{f(R)}) lead to an equivalent cosmological evolution to the general relativistic model filled with a fluid described  by  (\ref{EoS}). 
Therefore, as the LSBR is a future cosmic abrupt event of that model, then the reconstructed $f(R)$ theory will suffer the same classical fate.


\section{The LSBR in $f(R)$ quantum cosmology \label{sec:Qpart}}

 Despite of the lack of consensus about the existence a full quantum theory of gravity,  a quantum description of the Universe as a whole leads to an interesting framework, that is quantum cosmology
 (review on this topic can be found in references~\cite{Klaus libro,KlausBarbara}). Currently there are different approaches towards quantum cosmology. 
One of the first attempts to quantize cosmological backgrounds was due to DeWitt \cite{DeWitt}. 
In his work he provided a quantization procedure for a closed Friedmann Universe, leading to the first minisuperspace model in quantum cosmology  \cite{DeWitt, Kuchar}. 
The expression ``minisuperspace'' stands for a cosmological model truncated to a finite number of degrees of freedom. 
In addition, DeWitt proposed a criterion for the avoidance of classical singularities within this quantum framework. This is, the classical singularity is potentially avoided if the wave function of the Universe vanishes in the nearby configuration space. This criterion is, therefore, based on a probabilistic interpretation of the wave function, which would allow us to conclude that the probability to reach the singularity is zero. However, we have to stress that, unfortunately, this formulation is unknown in general.

In this section we will make use of the quantum geometrodynamics approach for the particular $f(R)$ theory we have obtained in the previous section. This approach is based on a canonical quantization with the Wheeler-DeWitt (WDW) equation playing a central role~\cite{DeWitt,KlausBarbara,Kuchar, Wheeler,Klaus libro, Intro QGeometro}. 
Then we will evaluate the quantum fate of the LSBR abrupt event with the DeWitt (DW) criterion. This criterion has been successfully applied in several cosmological scenarios in previous studies, e.g.  references~\cite{Klaus libro, fateQsingularities,Albarran:2016ewi, fRQC, EoS DE 2, Albarran,2016dcf,2018tel,2018mpg,2019zvz}.


\subsection{Modified Wheeler--DeWitt equation}\label{sec:MWD}

In cosmology, the gravitational action
\begin{equation}
S=\frac{1}{2}\int d^4x \sqrt{-g}f(R)  ,
\end{equation}
can be reformulated as
\begin{equation}
S=\frac{1}{2}\int dt \ \mathcal{L}(a,\dot{a},\ddot{a})  ,
\end{equation} 
taking the form of metric (\ref{metric}) into account.
In the preceding action, the Lagrangian is expressed by means of
\begin{equation}
\mathcal{L}(a,\dot{a},\ddot{a})=\mathcal{V}_{(3)}\,a^3f(R)  ,
\end{equation} 
with $\mathcal{V}_{(3)}$ the spatial $3$-dimensional volume. 
 As it was pointed in reference~\cite{Vilenkin}, for the canonical quantization of the theory a new variable can be introduced in order to remove the dependence on $\ddot{a}$ and to make clear the existence of an additional degree of freedom in metric $f(R)$ gravity.
It is useful to choose the scalar curvature to be the new variable, as in  references~\cite{fRQC, Vilenkin}. However, owing to the fact that  $R$ and $a$ are not independent
(their dependence is expressed in equation~(\ref{R(H)})), their relation needs to be introduced via a Lagrange multiplier $\mu$ for the constraint $R=R(a,\dot{a},\ddot{a})$. Thence, 
\begin{equation}
\mathcal{L}=\mathcal{V}_{(3)} a^3\left\lbrace f(R)-\mu\left[ R-6\left(\frac{\ddot{a}}{a}+\frac{\dot{a}^2}{a^2}+\frac{k}{a^2}\right) \right]\right\rbrace   .
\end{equation}
After solving for the Lagrange multiplier, the Lagrangian can be rewritten as~\cite{fRQC, Vilenkin}
\begin{equation}
\begin{split}\label{L}
&\mathcal{L}(a,\dot{a},R,\dot{R})=\mathcal{V}_{(3)}\left\lbrace a^3\left[f(R)-R f_R(R)\right]\right. \\
&\quad \left. -6a^2f_{RR}(R)\dot{a}\dot{R}+6af_R(R)(k-\dot{a}^2)\right\rbrace   ,
\end{split}
\end{equation}
with the notation $f_R\equiv df/dR$ and $f_{RR}\equiv d^2f/dR^2$. The derivative part of the Lagrangian  is not in a diagonal form, which leads to a quite unhandleable expression when  considering the quantization procedure. 
To overcome this issue we perform a change of variables alike to that described by Vilenkin in reference~\cite{Vilenkin}. That is
\begin{equation}\label{Vilenkin qx} 
q=a\sqrt{R_0}\left(\frac{f_R}{f_{R_0}}\right)^{\frac{1}{2}}\quad  \text{and} \quad
x=\frac{1}{2}\ln\left(\frac{f_R}{f_{R_0}}\right)   ,
\end{equation}
where $f_{R_0}\equiv f_R(R_0)$ and $R_0$ is a constant needed for the new variables to be well-defined. (We address further discussion on the value of $R_0$ to section \ref{sec:WDWfR}.) 
Consequently, the Lagrangian from (\ref{L}) becomes
\begin{equation}\label{L(x,q)}
\begin{split}
\mathcal{L}(x,\dot{x},q,\dot{q})&=\mathcal{V}_{(3)}\left(\frac{R_0f_R}{f_{R_0}}\right)^{-\frac{3}{2}}q^3 \left[f-6f_R\frac{\dot{q}^2}{q^2}\right. \\
&\quad \left. -Rf_R +6f_R\dot{x}^2+6k\frac{R_0}{f_{R_0}}\frac{f^2_R}{q^2}\right]   ,
\end{split}
\end{equation}
where $f$ and $f_R$ are now understood as functions of $x$.

Once the  derivative part has been diagonalized, we can proceed to obtain the corresponding Hamiltonian. The conjugate momenta are
\begin{align}
P_q&=\frac{\partial\mathcal{L}}{\partial \dot{q}}=-12\mathcal{V}_{(3)} R_0^{-\frac{3}{2}}f^{\frac{3}{2}}_{R_0}f^{-\frac{1}{2}}_R q \dot{q} \ \ ,\\
P_x&=\frac{\partial\mathcal{L}}{\partial \dot{x}}=12\mathcal{V}_{(3)} R_0^{-\frac{3}{2}}f^{\frac{3}{2}}_{R_0}f^{-\frac{1}{2}}_R q^3 \dot{x}   .
\end{align}
Therefore, the Hamiltonian reads
\begin{equation}\label{Hamiltonian}
\begin{split}
& \mathcal{H}=-\mathcal{V}_{(3)}q^3\left(\frac{R_0f_R}{f_{R0}}\right)^{-3/2} \left\{ f+6k\frac{R_0}{f_{R0}}\frac{f_R^2}{q^2} \right. \\
 &\quad \left. -Rf_R+
 \frac{6R_0^3}{(12)^2\mathcal{V}_{(3)}^2f_{R0}^3}\frac{f_R^2}{q^4}\left[P_q^2-\frac{P_x^2}{q^2}\right]\right\}  .
\end{split}
\end{equation}

For the quantization procedure, we assume $P_q\to-i\partial_q$ and $P_x\to -i\partial_x$. Then, the classical Hamiltonian constraint $\mathcal{H}=0$ becomes the modified WDW equation for the wave function $\Psi$ of the Universe \cite{Klaus libro,Vilenkin, DeWitt}. This is
\begin{equation}
\mathcal{\hat{H}}\Psi=0   .
\end{equation}
After some manipulations, the former expression can be rewritten as \cite{Vilenkin}
\begin{equation}\label{MWDW}
\left[q^2\partial^2_q-\partial^2_x-V(x,q)\right]\Psi(x,q)=0 ,
\end{equation}
where the effective potential is given by
\begin{equation}\label{V}
V(x,q)=\frac{q^4}{\lambda^2}\left(k+\frac{q^2}{6R_0f_{R_0}}(f-Rf_R)e^{-4x}\right)  ,
\end{equation}
with $\lambda=R_0/(12\mathcal{V}_{(3)}f_{R0})$. Note that when the expression of the $f(R)$ is given, the variables $x$ and $q$ in  (\ref{Vilenkin qx})  are completely set. Then, $f$ and $Rf_R$ must be expressed in terms of $x$.


\subsection{Quantum treatment of the LSBR\label{sec:WDWfR}}

Now, let us focus our attention on the particular expression for $f(R)$ given by the reconstruction method showed in section \ref{sec:fRLSBR}, this is equation  (\ref{f(R)}). 
Note that the term with $c_2$ cannot be directly  expressed through elemental functions of $R$. 
This feature prevents us from inverting the relations in equation~(\ref{Vilenkin qx}), i.~e.~from obtaining $R=R(x)$ in terms of elemental functions. 
 However, this is crucial for computing the WDW equation through  the path previously described. Therefore, for the sake of simplicity, we set $c_2=0$ to consider the study of a simple, still general, $f(R)$ cosmological model with a LSBR. 
 This model is given by
\begin{equation}\label{f(R)c2=0}
f(R)=c_1 \left(A^2-6AR+R^2\right)   .
\end{equation}
For this model, the change of variables (\ref{Vilenkin qx}) reads
\begin{equation}
q=a\sqrt{R_0}\left(\frac{R-3A}{R_0-3A}\right)^{\frac{1}{2}}   , \ x=\frac{1}{2}\ln\left(\frac{R-3A}{R_0-3A}\right). 
\end{equation}
Regarding the value of $R_0$, in  reference~\cite{Vilenkin} the curvature of the self-consistent de Sitter solution was proposed as a possible preferred value. In that case $R_0$ would stand for the solution to $R_0f_{R_0}-2f(R_0)=0$.
Nevertheless, this choice may not always be convenient, as it was shown in  reference~\cite{fRQC}.
In our case, if we adopt the definition through the de Sitter solution we would obtain $R_0=A/3$ and, therefore, $R_0-3A<0$, changing sign as $R$ increases. Therefore, this choice is not compatible with a well-defined change of variables given by (\ref{Vilenkin qx}).
On the other hand, note that (\ref{rho(a)}) and (\ref{R(H)})  imply 
\begin{equation}
R=4\Lambda+A\,\left[1+\ln (a/a_0)^4\right].
\end{equation}
Thus, following Vilenkin's spirit for a physical meaningful constant $R_0$, we define $R_0 = 4\Lambda+A$, which corresponds to the present value of the scalar curvature. As $A$ is small, we ensure $R_0-3A=4\Lambda-2A>0$ $(2\Lambda >A)$. Thus, as for our model $R$ is an increasing function in the future, the change of variables given by (\ref{Vilenkin qx}) is suitable to study the cosmic future.

A straightforward substitution of equation (\ref{f(R)c2=0}) in equations (\ref{MWDW}) and (\ref{V}) leads to the modified WDW equation for our model
\begin{widetext}
\begin{equation}\label{WDW fR}
\left\{q^2\partial_q^2-\partial_{x}^2+\frac{q^6}{12\lambda^2 R_0(R_0-3A)}\left[8A^2 e^{-4x}+6A(R_0-3A)e^{-2x}+(R_0-3A)^2 \right]\right\}\Psi(q,\,x)=0  \,,
\end{equation}
\end{widetext}
where we have assumed a spatially flat Universe, that is $k=0$.

As the main motivation of the present work is the evaluation of the wave function $\Psi$ at the LSBR regime, it is not necessary to find the whole solution to the WDW equation in the complete configuration space but only in the region close to the LSBR abrupt event. The most important condition for the occurrence of this doomsday is the divergence of the scalar curvature $R$ at an infinite cosmic time, which corresponds to $x\to\infty$ and $q\to\infty$. In addition, given that we are mainly interested in the asymptotic behaviour of the wave function $\Psi$, further simplifications can be made. 
Note that for $x\to \infty$, 
\begin{align}
8A^2e^{-4x}&\ll (R_0-3A)^2  , \\
6A(R_0-3A)e^{-2x}&\ll(R_0-3A)^2  .
\end{align} 
Consequently, in the region close to the LSBR abrupt event the potential dominant term depends only on one of the variables, that is 
\begin{equation}
V(x,q)\sim\frac{1}{12\lambda^2R_0}(R_0-3A)\,q^6  .
\end{equation}
Hence, the modified WDW equation is reassembled as 
\begin{equation}\label{WDW LSBR}
q^2\partial^2_q \Psi-\partial^2_x\Psi+Bq^6 \Psi=0  ,
\end{equation}
where, for the sake of clarity, we have defined
$B=\frac{1}{12\lambda^2R_0}(R_0-3A)$.
The solution of this equation can be found with the ansatz for the wave function 
\begin{equation}
\Psi (x,q)=\sum_{\tilde{k}} b_{\tilde{k}}C_{\tilde{k}}(x) U_{\tilde{k}}(q)  ,
\end{equation}
where $b_{\tilde{k}}$ stands for the amplitude of each solution and $\tilde{k}$ is related with the associated energy. Do not confuse $\tilde{k}$ with the spatial curvature $k$, which has been set to zero since the spatial curvature term is subdominant close to the LSBR. As a result, the WDW equation in (\ref{WDW LSBR}) implies the following equations
\begin{align}
\frac{d^2}{dx^2} C_{\tilde{k}}(x)-{\tilde{k}}^2C_{\tilde{k}}(x)&=0   , \\
\label{eqU}
\frac{d^2}{dq^2}U_{\tilde{k}}(q)+ \left(Bq^4-\frac{\tilde{k}^2}{q^2} \right)U_{\tilde{k}}(q)&=0  .
\end{align}
The first equation can be directly solved
\begin{subequations}\label{solC}
\begin{align}
C_{\tilde{k}}(x)=a_1e^{\tilde{k}x}+a_2e^{-\tilde{k}x} \ \ &\text{for} \ \tilde{k}^2\geq 0   , \label{solCk2+} \\
C_{\tilde{k}}(x)=a_3e^{i|\tilde{k}|x}+a_4e^{-i|\tilde{k}|x} \ \ &\text{for} \ \tilde{k}^2<0 \label{solck2-}  ,
\end{align}
\end{subequations}
being $a_1$, $a_2$, $a_3$ and $a_4$  arbitrary constants.
 
On the other hand, the equation for $U_{\tilde{k}}(q)$ admits an exact solution when $\tilde{k}^2=0$ by means of Bessel functions, cf.~equation~9.1.51 of reference~\cite{libroFunciones},
\begin{equation}\label{U0}
U_0(q)=\sqrt{q} \left[d_1 J_{\frac{1}{6}}\left(\frac{\sqrt{B}}{3} q^3\right)+d_2 Y_{\frac{1}{6}}\left(\frac{\sqrt{B}}{3} q^3\right) \right]   ,
\end{equation}
being $J_{\frac{1}{6}}$ and $Y_{\frac{1}{6}}$ the Bessel functions of first and second kind, respectively, and $d_1$ and $d_2$ constant parameters. 
For values of $\tilde{k}^2\neq 0$ the solution can be approximated making use of the WKB method. In Appendix \ref{WKBapp}, we found that the first order WKB approximation leads to
\begin{equation}\label{Uk}
U_{\tilde{k}}(q)=\left(Bq^4-\frac{\tilde{k}^2}{q^2}\right)^{-\frac{1}{4}}\left[d_3e^{i I} +d_4e^{-iI}  \right],
\end{equation}
where $I$ is defined by
\begin{equation}\label{Ik<0}
\begin{split}
I&=\frac{1}{3}\sqrt{Bq^6+|{\tilde{k}}^2|} \\
&\quad  -\frac{\sqrt{|{\tilde{k}}^2|}}{3}\coth^{-1} \left(\sqrt{\frac{B}{|{\tilde{k}}^2|}q^6+1} \right) +C,
\end{split}
\end{equation}
for $\tilde k^2<0$
or
\begin{equation}\label{Ik>0}
\begin{split}
I&= \frac{1}{3}\sqrt{Bq^6-{\tilde{k}}^2}\\
&\quad +\frac{\sqrt{{\tilde{k}}^2}}{3}\cot^{-1} \left(\sqrt{\frac{B}{{\tilde{k}}^2}q^6-1} \right)+C,
\end{split}
\end{equation}
for $\tilde{k}^2>0$ where $C$ is an arbitrary constant. The solutions (\ref{U0}) and (\ref{Uk}) exhibit all the same asymptotic behaviour, this is
\begin{equation}
U_{\tilde{k}}(q)\sim \frac{1}{q} \left[u_1 \cos (\frac{\sqrt{B}}{3}q^3)+u_2 \sin (\frac{\sqrt{B}}{3}q^3)\right]  ,
\end{equation}
whatever the value of $\tilde{k}^2$. Note that $u_1$ and $u_2$ are integration constants.
So, all $U_{\tilde{k}}(q)\rightarrow0$ when $q\rightarrow\infty$.

In summary, the solution for $C_{\tilde{k}}(x)$  with $\tilde{k}^2<0$ are finite for any value of $x$. Nevertheless, equation (\ref{solCk2+}) remains bounded if and only if  $a_1=0$.
Therefore, the choice $a_1=0$ leads to a finite solution for $C_{\tilde{k}}(x)$. 
In addition, the solution for $U_{\tilde{k}}(q)$ shrinks to zero as $q$ tends towards infinity. Hence, for the choice $a_1=0$, the wave function $\Psi (x,q)$ vanishes at the LSBR regime. 
Thus, the DW criterion is satisfied, pointing towards the avoidance of the LSBR abrupt event in the quantum realm.


\section{Conclusions\label{sec:Summary}}
The LSBR is a cosmic abrupt event predicted in general relativistic phantom models with an equation of state that slightly depart from a cosmological constant. 
Considering the quantum cosmological framework based on quantum geometrodynamics, it has been shown that this event may be avoided when the corresponding classical cosmic evolution is described by GR \cite{Albarran,EoS DE 2}. In this work we have analysed whether this is still the case when the classical cosmic evolution is due to a $f(R)$ theory of gravity instead of a dark fluid.

Therefore, in the first place, we have obtained the group of metric $f(R)$ theories of gravity that predicts a LSBR abrupt cosmic event. 
We have used a reconstruction method to obtain the group of $f(R)$ theories able to mimic this particular cosmic evolution, which in GR corresponds to a phantom energy model.

In the second part of the work, we have investigated the quantum fate of the LSBR predicted by one of the obtained $f(R)$ theories of gravity. So, we have considered
the formulation of $f(R)$ quantum cosmology in the framework of quantum geometrodynamics for that particular theory. 
We have found the solutions of the modified WDW equation and show that those solutions satisfy the DW condition when one of the integration constant is set to zero. 
This fact points towards the avoidance of the LSBR abrupt event in $f(R)$ theories of gravity, since the wave function of the Universe vanishes at the corresponding point in the minisuperspace. 

It should be noted that, when applying the DW criterion, we have fixed to zero an integration constant, discarding a subgroup of solutions as unphysical. If future investigations leads to the need of taking into account the solution dismissed, then it would be concluded that the DW criterion may not always be satisfied.


\begin{acknowledgments}
MBL is supported by the Basque Foundation of Science Ikerbasque. She also would like to acknowledge the partial support
from the Basque government Grant No. IT956-16 (Spain) and from the project FIS2017-85076-P (MINECO/AEI/FEDER, UE).
PMM acknowledges financial support from the project FIS2016-78859-P (AEI/FEDER, UE).
\end{acknowledgments}


\appendix

\section{The WKB approximation}\label{WKBapp}
 
For a second order homogeneous ordinary differential equations of the form
\begin{equation}
 \epsilon^2\frac{d^2U}{dq^2}+Q(q)\,U=0 \ \ ,
\label{eqWKBejemplo}
\end{equation}
the unknown exact solution can be approximated to an exponential solution of the form  \cite{WKB book, WKB book 2, WKB book 3, WKB book 4}
\begin{equation}
 U(q)=\exp \left[{\frac {1}{\delta }}\sum _{n=0}^{\infty }\delta ^{n}S_{n}(q)\right].
\end{equation}
  Then, the first order WKB approximation reads
\begin{equation}
\begin{split}
U(q)&\sim Q^{-\frac{1}{4}} \left[\tilde{u}_1\exp\left(i\frac{1}{\epsilon}\int_{q_0}^q\sqrt{Q(z)}dz\right)\right.\\
&\quad \left.+\tilde{u}_2\exp\left(-i\frac{1}{\epsilon}\int_{q_0}^q\sqrt{Q(z)}dz\right)\right],
\end{split}
\end{equation}
where  $\tilde{u}_1$ and $\tilde{u}_2$ are constants to be determined from initial or boundary conditions and  $q_0$ is an arbitrary but fixed integration point.

In the case of (\ref{eqU}),  $Q(q)/\epsilon^2={Bq^4-\frac{{\tilde{k}}^2}{q^2}}$ with ${\tilde{k}}\neq 0$. 
Therefore, we obtain
\begin{equation}
U_{\tilde{k}}(q)\sim \left({Bq^4-\frac{{\tilde{k}}^2}{q^2}}\right)^{-\frac{1}{4}} \left[\tilde{u}_1e^{\left(iI\right)}+\tilde{u}_2e^{\left(-iI\right)}\right],
\end{equation}
with
\begin{equation}
\begin{split}
I&= \left[\frac{1}{3}\sqrt{Bq^6+|{\tilde{k}}^2|}\right. \\
&\quad \left. -\frac{\sqrt{|{\tilde{k}}^2|}}{3}\coth^{-1} \left(\sqrt{\frac{B}{|{\tilde{k}}^2|}q^6+1} \right)\right]_{q_0}^q   ,
\end{split}
\end{equation}
for $\tilde{k}^2<0$, and
\begin{equation}
\begin{split}
I&= \left[\frac{1}{3}\sqrt{Bq^6-{\tilde{k}}^2}\right. \\
&\quad \left.+\frac{\sqrt{{\tilde{k}}^2}}{3}\cot^{-1} \left(\sqrt{\frac{B}{{\tilde{k}}^2}q^6-1} \right)\right]_{q_0}^q   ,
\end{split}
\end{equation}
for $\tilde{k}^2>0$. 
It is worth to mention that the freedom of fixing the integration point $q_0$ can be used in such a way that $Q(q)\geq 0$ in the interval of integration. Consequently, $I$ is always real.

The validity of the first order WKB approximation is given by the fulfilling of the inequality \cite{ WKB book 2, WKB book 3, WKB book 4}
\begin{equation}
\left| \frac{Q'}{Q^\frac{3}{2}}\right|\ll 1 .
\end{equation}
In our case this leads to
\begin{equation}
\left| \left(4Bq^3+\frac{2\tilde{k}^2}{q^3}\right)\left(Bq^4-\frac{\tilde{k}^2}{q^2}\right)^{-\frac{3}{2}}\right|\ll 1,
\end{equation}
which is true for large values of $q$. Therefore, we conclude that the first order WKB approximation  for $U_{\tilde{k}}$ is valid in the region close to the LSBR abrupt event.


\end{document}